\documentclass[entropy,article,submit,oneauthor,dvipdfm,12pt,a4paper]{mdpi} 
\usepackage{epsfig,graphicx}
\usepackage{epstopdf}
\lastpage{x}
\doinum{10.3390/------}
\pubvolume{xx}
\pubyear{2013}
\history{Received: xx / Accepted: xx / Published: xx}

\Title{Complexity in Animal Communication: Estimating the Size of N-Gram Structures}

\Author{Reginald Smith $^{1,}$*}

\address{$^{1}$ Citizen Scientists League, PO Box 10051, Rochester, NY 14610, USA}

\corres{rsmith@citizenscientistsleague.com}

\abstract{In this paper, new techniques that allow conditional entropy to estimate the combinatorics of symbols are applied to animal communication studies to estimate the communication's repertoire size. By using the conditional entropy estimates at multiple orders, the paper estimates the total repertoire sizes for animal communication across bottlenose dolphins, humpback whales, and several species of birds for N-grams length one to three. In addition to discussing the impact of this method on studies of animal communication complexity, the reliability of these estimates is compared to other methods through simulation. While entropy does undercount the total repertoire size due to rare N-grams, it gives a more accurate picture of the most frequently used repertoire than just repertoire size alone.}


\begin{document}
\section{Introduction}
The complexity of animal communication is a topic frequently discussed, but difficult to resolve. While it is beyond dispute that many species communicate, even the basic purposes of these communications--whether to communicate information or to just influence the behavior of others to increase their own fitness--is hotly debated\cite{dissident1,dissident1b,dissident2,dissident3, rethink}. Even if we conclude information is being communicated, does the faculty for language, the human ability to communicate complex information through spoken language, have wide and directly comparable analogs across the animal kingdom \cite{pinkerjack} or is the faculty for language and expressing abstract ideas uniquely human \cite{hauserchomsky}?

The complexity of animal language has been studied using many methods including various techniques to estimate repertoire size such as curve-fitting \cite{wildenthal, catchpole} and capture-recapture \cite{catchpole, garamszegi1,garamszegi2, botero}. Other methods use information theory either by measurements of conditional entropy \cite{lemondobson,dolphin1} or using other methods such as entropy rate and Lempel-Ziv complexity \cite{kershenbaum}. In this paper, we will focus on the methods using conditional entropy. Measuring animal communication in terms of the entropy in bits, these studies have attempted to look at the animal communication structure at various lengths (N-grams) in order to determine the structure of the communications and whether the tools of information theory can lend themselves to a better understanding of animal behavior and possibly what types of information can be communicated.

\section{Information Theory and Animal Communication}

After formulating information theory in 1948, Shannon was not long in turning its powers to shedding light on human language \cite{shannonenglish}. Shannon investigated the entropy of the English language using both frequency counts of letters from texts as well as human volunteers who played a guessing game of missing letters to establish bounds of the estimated entropy. This analysis of language mainly focused on the measure of what is now widely known as the conditional entropy. The conditional entropy of order $N$ is defined with the probability of a given letter ($j$) coming after an $N$-gram sequence ($b_i$).

\begin{equation}
H_N = -\sum_{i,j}p(b_i, j) \log_2 p_{b_i}(j)
\label{condent}
\end{equation}

Where $p(b_i, j)$ is the joint probability of the sequence ($b_i,j$) and $p_{b_i}(j)$ is the conditional probability of $j$ given $b_i$. The conditional entropy for $N=2$ is often written as $H(X|Y)$ and can have a maximum value of $H(X)$. For $N$=1 this reduces to the well-known Shannon entropy.

\begin{equation}
H = -\sum_{i=1}^M p_i \log_2 p_i
\end{equation}

Amongst the simplest methods for computing conditional entropies is from joint entropies. The joint entropy, $H(N)$, for a sequence of symbols ($x_i$) of length $N$ is defined as 

\begin{equation}
H(N) = -\sum_{x_1}\dots \sum_{x_n} p(x_1,\dots,x_n) \log_2 p(x_1,\dots,x_n)
\end{equation}

The conditional entropy of order $N$ can be alternatively defined as $H_N = H(N) - H(N-1)$ where $H(N)$ and $H(N-1)$ are the joint entropies of order $N$ and $N-1$ respectively.

For the English alphabet of 27 letters (26 letters plus the space character), Shannon calculated the first order entropy at 4.14 bits, the second order conditional entropy at 3.56 bits, and the third order conditional entropy at 3.30 bits. The zero-th order entropy of 4.75 bits was based on $\log_2 M$ where $M$=27. Many other languages have been analyzed in this way across many language families. Data and analysis for a large group of these are given in \cite{yaglom, Smith}.

Soon after human languages, animal communication of varying types were studied using entropy. One of the first citations explicitly analyzing animal communication by means of information theory was that of J.B.S. Haldane and H. Spurway \cite{haldane} who did a short calculation to estimate the information entropy of bee (\emph{Apis Mellifera}) dances at 2.54 bits. Many modern treatments of animal communication by information theory can be traced to the work of Chatfield \& Lemon on cardinals (\emph{Cardinalis cardinalis}) \cite{lemon1,lemon2} and Lemon \& Dobson on thrushes \cite{lemondobson}. In particular, their work on analyzing different orders of entropy to investigate the fundamental order of communication established a baseline on using information theory to estimate the complexity of animal communication.

Further studies along this line include the analysis of the chickadee (\emph{Parus atricapillus}) and (\emph{P. carolinensis}) by \cite{chick1,chick2}, European starlings (\emph{Sturnus vulgaris}) \cite{starling}, Rufous bellied thrushes (\emph{Turdus rufiventris}) \cite{thrush}, European skylarks (\emph{Alauda arvensis L.}) \cite{skylark}, wood thrushes (\emph{Hylocichla mustelina}) and robins (\emph{Turdus migratorius}) \cite{lemondobson}, bottlenose dolphins (\emph{Tursiops truncatus}) \cite{dolphincond,dolphin1,dolphin2,dolphin3}, humpback whales (\emph{Megaptera novaeangliae}) \cite{whale1,whale2,whale3,whalecond}, and male rock hyraxes (\emph{Procavia capensis}) \cite{hyraxes}. 

These studies are primarily focused on measuring information through entropy in bits in the first order, and sometimes higher orders as well. For multiple orders, information graphs, plots of the bits of conditional entropy by order, are sometimes used \cite{lemondobson} to analyze the structure of the communication and estimate the Markov order of the signal. While this provides a quantitative overall measure of complexity, they have a limitation in that they do not provide resolution into how many, or what type, of calls or songs that we should expect in two, three or more combined units. Using the values of entropy, few conclusions can be deduced besides the order at which a signal becomes most repetitive: where its value drops most sharply from one order to the next. To remedy this, we can use information theory with combinatorics so that the size of the repertoire, at lengths longer than one, can be estimated with only information about the conditional entropy for each order.

\subsection{Information Graphs and Order Complexity}

An information graph is the plot of the higher order conditional entropies by order. Some of the first uses and analyses of information graphs in the context of Markov sequences are given in \cite{chatfield3, morgan}. Information graphs were first used to analyze the order dependence of Markov sequences, the theory being that when there is a large, negative slope between two orders to a relatively low value of conditional entropy, the prior order is most likely the order of dependence of the Markov sequence to describe the communication. However, \cite{morgan} showed through simulation that a large decrease between two orders of entropy in an information graph cannot be determined to be the fundamental order if the number of symbols is high or the sample size is low. Since likelihood tests become unreliable at smaller sample sizes with large symbol alphabets, the decrease in the information graph could be indicative of the inadequacy of sample sizes at larger orders rather than the fundamental order of the underlying Markov process.

With these caveats, the information graphs will still be shown as an illustration of the results of the studies on each animal communication and should be used with caution to establish the complexity of sequences.

\begin{figure}
\centering
 \begin{tabular}{cc}
 	 \includegraphics[height=3in, width=3in]{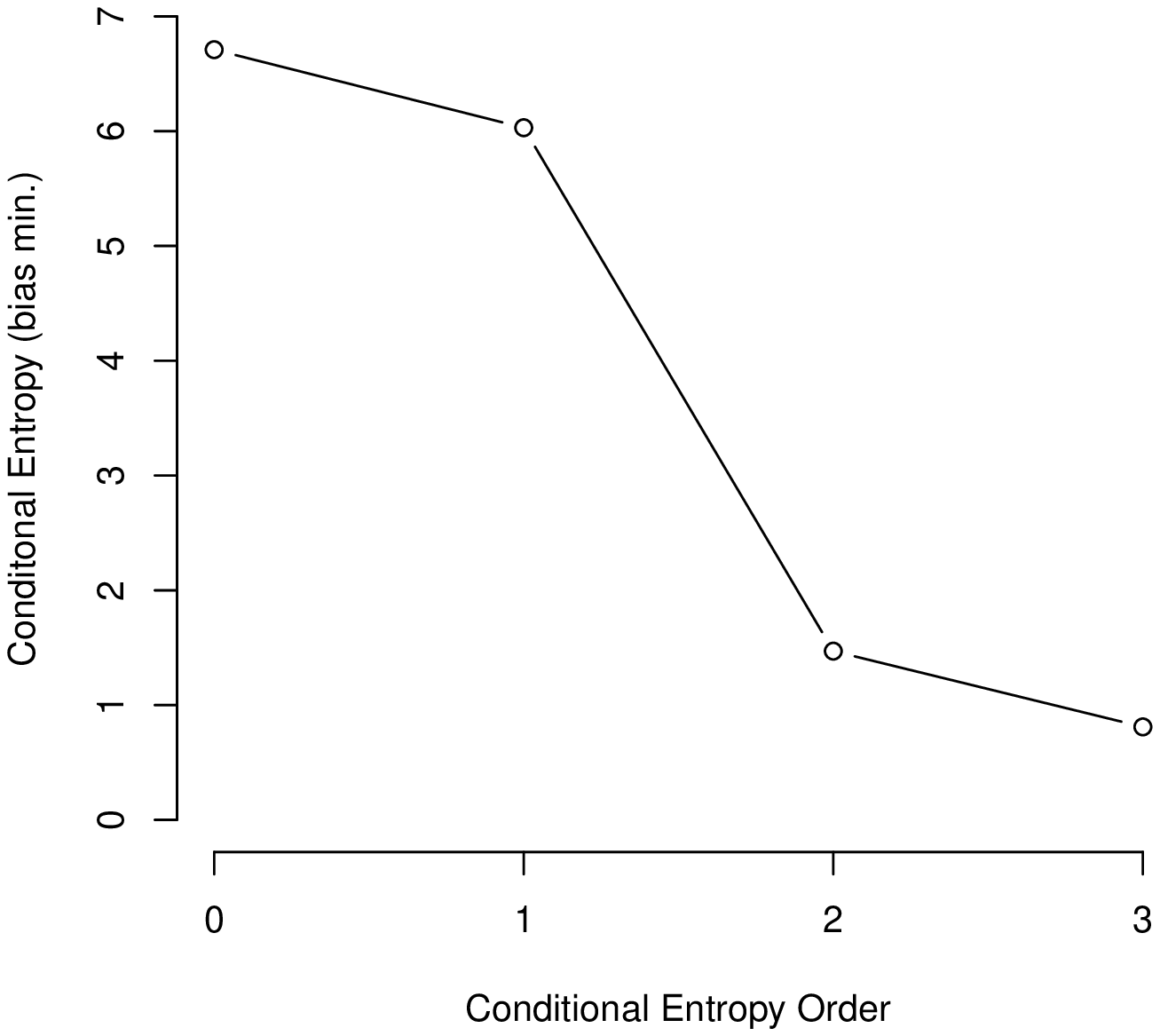}&
    \includegraphics[height=3in, width=3in]{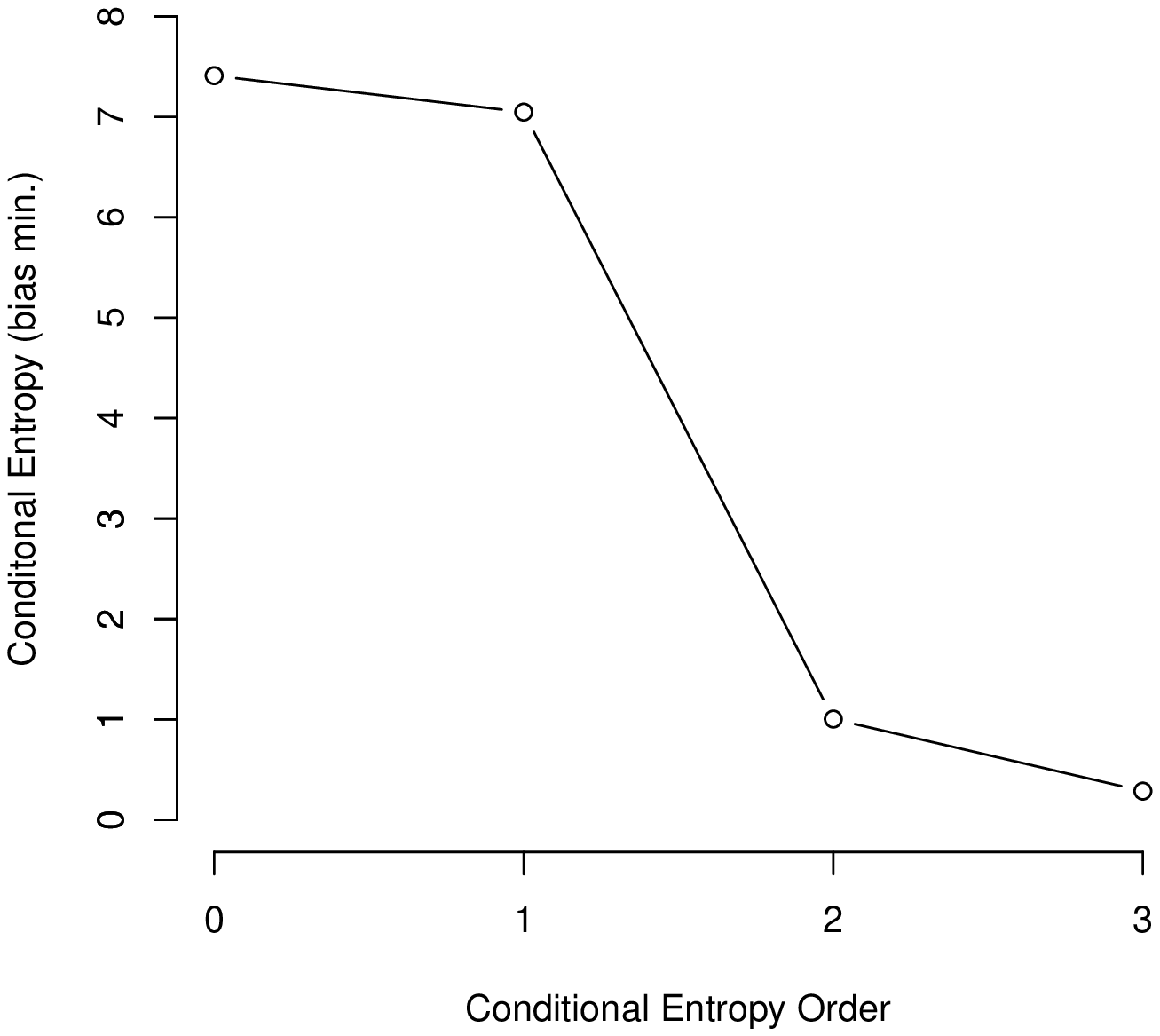}\\
 \small{European skylarks (\emph{Alauda arvensis L.})}&\small{European starlings (\emph{Sturnus vulgaris})}\\
\end{tabular}
\caption{Information graphs of communications by European skylarks \cite{skylark} and European starlings \cite{starling} adjusted for minimum bias (see Table \ref{correntropytable}).}
\label{birdinfo}
\end{figure}

In general, the larger the order of dependence, the more ``complex'' the communication is deemed. For example, many bird call sequences seem to show first order dependence, though this is unsure since a sample size of multiples of the number of symbols squared is needed to confirm this (Figure \ref{birdinfo}). This is much different than human written language. In a point first made in \cite{chick1}, English written letters show a drop of less than 1 bit from the first to third order conditional entropies \cite{shannonenglish, dolphincond}, much slower than the drop in the chickadee information graphs and those of other birds. 

While information graphs are relatively easy to construct given the right data, there is a large issue of estimating entropy. Namely, entropy estimators can have large biases, that depend on the sample size, which typically underestimate the true value of entropy \cite{miller, panzeri}.


\subsection{Bias Measures in Entropy Estimates}

Because of the often large numbers of possible variables, entropy estimators can be very sensitive to sample size and introduce bias into measurements. This was first investigated in \cite{miller} where the following expression is the first order additive term to entropy estimates to correct for bias

\begin{equation}
H = \mbox{\^{H}} + \frac{M-1}{2S}
\label{bias1}
\end{equation}

where, $\mbox{\^{H}}$ is the entropy estimator based on the data, $M$ is the number of non-zero categories across which the probabilities are measured to calculate entropy and $S$ is the sample size. This estimator was improved in \cite{panzeri} as

\begin{equation}
H = \mbox{\^{H}} + \frac{M-1}{2S \ln{2}}
\label{bias2}
\end{equation}

When dealing with actual data, it can be relatively straightforward to estimate $M$, though with smaller sample sizes it is questionable if you have captured all non-zero categories. However, when only sample sizes and values for entropies are available, calculating $M$ accurately can be much more difficult. With little information available, we can estimate upper and lower bounds for the entropy bias. This will be described following the section on combinatorics.

\section{Combinatorics of Information Theory and Repertoire Size}

One of the lesser known, but extremely useful, facets of information theory is the way entropy can be used for combinatorics. In particular, the number of  combinations of a symbol set can be more accurately estimated using the first-order entropy than can be done with an assumption of random likelihood. For example, if an alphabet has $M$ symbols, the exact number of possible combinations of length $N$ is the common result

\begin{equation}
W_N = M^N
\label{combo1}
\end{equation}

Here, $W_N$ is the total number of possible combinations of length $N$. This basic calculation assumes every combination appears with non-zero probability. This can be improved on, however, using the calculation from Shannon and Weaver \cite{weaver} if we know the first order entropy. Here we can estimate the number of combinations that appear with probability 1 assuming the measurement of first order entropy is accurate:

\begin{equation}
W_N = M^{NH_{(\log M)}}
\label{shannonweaver1}
\end{equation}

Here $H$ is the Shannon (first-order) entropy using logarithm of base $M$. This assumes that each symbol in the N-Gram appears with a rate based on the entropy of the symbol alphabet. Clearly, if each symbol is equally likely, $H$ is at most 1 and we get Equation (\ref{combo1}). The more familiar version (the one derived by Shannon and Weaver) calculates $W_N$ using entropy in units of bits (log base 2)

\begin{equation}
W_N = 2^{NH_{(\log 2)}}
\label{shannonweaver2}
\end{equation}

Equations (\ref{shannonweaver1}) and (\ref{shannonweaver2}) improve on the assumptions of Equation (\ref{combo1}) by incorporating the fact that every symbol is not equally likely but appears at a rate consummate with the Shannon entropy of the overall signal. These derivations show that knowledge of the entropy of the signal allows us to reduce the number of combinations and more accurately estimate the number of combinations of length $N$. However, there is an additional element of error in this analysis.

Since $H$ is the first-order entropy, this Shannon-Weaver model assumes that each symbol has an i.i.d. probability of appearing in each space in the N-Gram. If there is any correlation between symbols, the larger $N$ becomes, the more likely $W_N$ is inaccurate. However, in this model there is no co-dependence between symbols on which symbol is more likely to follow another and the base assumption is that in any $N$ length string, the symbols for each position are chosen independent of all other symbols before them.

In order to improve on the estimate of $W_N$ for $N > 1$, we must use the conditional entropy. In a result first demonstrated by Kolmogorov \cite{kolmogorov}, $W_N$ can be more accurately estimated by using conditional entropy to account for all possible pairs, without the overlap instances that are found in the Cartesian product (represented by joint entropy) of the alphabet spaces. Note that in his paper, Kolmogorov stated that $W_2=2^{H(X|Y)}$ However, a factor of two is necessary for the equation to reduce to the base case of Shannon and Weaver if $H(X|Y)=H(X)$.

\begin{equation}
W_2 = 2^{2H(X|Y)}
\label{kolmo}
\end{equation}

In the above $H(X|Y)$, also expressed as $H_2$, is the conditional entropy in bits for the digram sequence $XY$. Given the inequality $H(X) \geq H(X|Y)$, Equation (\ref{kolmo}) reduces to Equation (\ref{shannonweaver2}) at maximum conditional entropy where co-dependence disappears. Equation (\ref{kolmo}) was originally used to calculate the number of digrams but can extended for $N > 2$ using higher order conditional entropies. If we designate conditional entropies of order $N$ as $H_N$ the upper bound estimate of the number of combinations of length $L$, $W_L$, where $N \leq L$ is

\begin{equation}
W_L = 2^{LH_N}
\end{equation}

Since conditional entropy must monotonically decrease with each higher order, $W_L$ is at a minimum where $N=L$ since $H_L$ is smaller than all preceding conditional entropies. This can apply to language in some obvious ways. For example, an estimate of the number of distinct two-letter words in a language can be given by $W_2=2^{2H_2}$. For distinct three-letter words we can use $W_3=2^{3H_3}$ etc. This approach, along with a new statistical distributional approach, was demonstrated in \cite{Smith}. Using these parameters then, it is an intriguing question if we can estimate the size of the repertoire of multiple symbols or sounds in non-human systems of communication.

\subsection{Combinatorics and Entropy Bias Estimates}

In addition to estimating the size of the repertoire, combinatorics can be used to estimate upper bounds for the entropy bias when details about the data set are unavailable. This is primarily through estimating $M$, the number of non-zero categories in Equations (\ref{bias1}) and (\ref{bias2}). The upper bound for $M$, given a specific order of entropy, $H$, can be estimated using the assumptions of Equation (\ref{shannonweaver2}). The largest possible value for $M$ for an order, $N$, of entropy $H$ can be given by $M=2^{NH}$. Therefore, if the bias of $H$ using the number of symbols is acceptably low, $M=2^{NH}$ can be used in calculations to find the largest possible bias expected for a given sample size. 

In addition, one can estimate a lower bound for $M$ using the combinatorics of conditional entropies. The lower bound for $M$ should be $M=2^{NH_N}$. With these two values of $M$, we can determine an appropriate band for the repertoire for any order. The largest problem can occur if $H$ is relatively large with a low order of dependence. This can make the upper bound estimation of bias huge, with the lower bound relatively small. As will be seen later, this can be an issue with birds with a large repertoire of individual calls but with a relatively low (second order) dependence in their communication. As a final note, the bias corrections apply only to the first, second, and third order \emph{joint} entropies. These are then subtracted from one another to find the bias corrected conditional entropies.

In the next section, we will investigate the complexity of several species including bottlenose dolphins, humpback whales, and several species of birds and investigate the size of their N-gram repertoires.

\section{Animal Communication: Complexity and Repertoire Size}

In this paper we will use entropy combinatorial techniques to estimate the N-gram repertoires of six species: bottlenose dolphins \emph{Tursiops truncatus} \cite{dolphincond,dolphin1,dolphin2,dolphin3}, humpback whales \emph{Megaptera novaeangliae} \cite{whale1,whale2,whale3,whalecond}, European starlings \emph{Sturnus vulgaris} \cite{starling}, European skylarks \emph{Alauda arvensis L.} \cite{skylark}, wood thrushes \emph{Hylocichla mustelina} and robins \emph{Turdus migratorius} \cite{thrush}. A brief summary of the research for each is given below, followed by data from the papers, information graphs, and estimated N-gram repertoire sizing.

\subsection{Bottlenose Dolphins}

In \cite{dolphinwhistle1,dolphinwhistle2,dolphinwhistle3} McCowan and Reiss introduced a new method to categorize the whistles of bottlenose dolphins,  \emph{Tursiops truncatus}, and organize these into sequences. This research was followed up in a collaboration with Doyle \cite{dolphincond} which analyzed these sequences in terms of information theory and Zipf's Law calculating the conditional entropy up to order three, comparing this with human written language, and calculating a Zipf exponent of nearly -1 for the rank-frequency distribution of dolphin whistle types. This paper will use the data from \cite{dolphincond} to investigate the dolphin whistles for N-grams for $N$ in range one to three.

\subsection{Humpback Whales} 

One of the defining features of humpback whales, \emph{Megaptera novaeangliae}, is their social organization into groups called pods where they emit various cries, both alone and in sequence, to communicate with other whales. These cries were investigated through the lens of information theory in several papers \cite{whale1,whale2,whale3,whalecond}. Suzuki \cite{whale1} and Miksis-Olds and collaborators \cite{whale2,whale3} analyzed the structure of humpback  whale mating songs and found both that the sequences of whale cries were not stationary and could not be represented well by a first order Markov chain model. Doyle and collaborators in \cite{whalecond} investigate the entropy and conditional entropy of humpback whale cries under conditions of man-made noise and relative quiescence in order to establish how anthropomorphic noise may affect whale cry patterns. They found a significant effect where whale cries seemed to have a steeper entropic slope, and are thus more repetitive, under high noise conditions, possibly to compensate for the more noisy channel. For our analysis, we will use the results from the low noise data set.

\subsection{Wood Thrushes and Robins}

Dobson and Lemon \cite{lemondobson} investigated the information structure of long call sequences amongst a variety of American thrushes including wood thrushes \emph{Hylocichla mustelina} and robins \emph{Turdus migratorius}. For each bird they measured multiple sequences and calculated entropies of the call sequences to create information graphs. Being one of the earliest papers to use this technique on animal communication, it established many methods such as the use of information graphs. In this paper, we will look at the entropies based on the subjects of the paper, wood thrush 3 and robin 2.

\subsection{European Skylarks}

In \cite{skylark}, Briefer and collaborators measured the information entropy of European skylarks in both France and Poland to test the hypothesis that habitat change, marked in France but not Poland, is having a significant effect on the call patterns of \emph{Alauda arvensis L.}. While songs were more shared amongst different birds in the restricted habitat near Paris, song complexity was almost identical in both locations. For this paper, we use the continuous habitat data from the Poland habitat.

\subsection{European Starlings}

In \cite{starling}, Getner and Hulse investigated the ability of European starlings, \emph{Sturnus vulgaris}, to recognize individuals based on songs. As part of their analysis, they used a success-failure reward to access a food hopper based on correctly distinguishing one starling call amongst a group of five. When they used synthetic call sequences to test recognition, they found recognition was improved when sequences with second or third order Markov dependence (more complex) were used versus first order dependencies which randomly emitted sounds with a frequency to match first-order entropy. For this paper, we will use the data from the entropy of song types in starling song bouts represented in an information graph in the paper's Figure 3. Since the sample size was not explicitly mentioned in the paper, it was estimated by using data from the paper. Namely, assuming a song type (syllable) average length of one second, an average of about 39s per song bout, and 120 song bouts. This gives $S$=4,680. In addition, since each bout had a standard error of 6s, we used the $2*SE$ 95\% confidence interval to add an additional $2*SE*\sqrt{120}$ seconds for a total sample time (and sample size) of 4,811.

\section{Animal Communication Entropy Data and Repertoire Estimates}

\begin{figure}
\centering
 \begin{tabular}{cc}
	 \includegraphics[height=2.5in, width=2.5in]{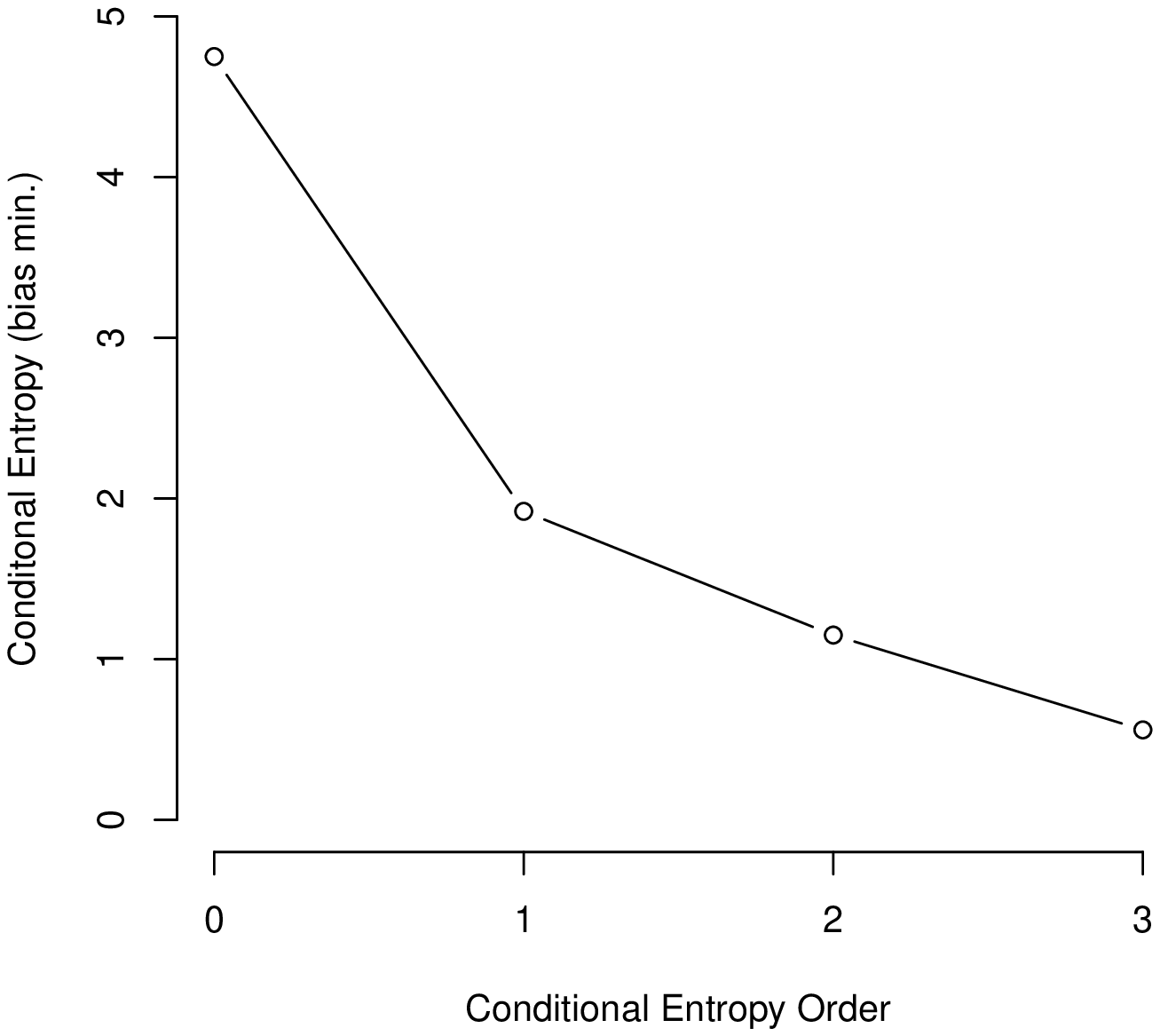}&
    \includegraphics[height=2.5in, width=2.5in]{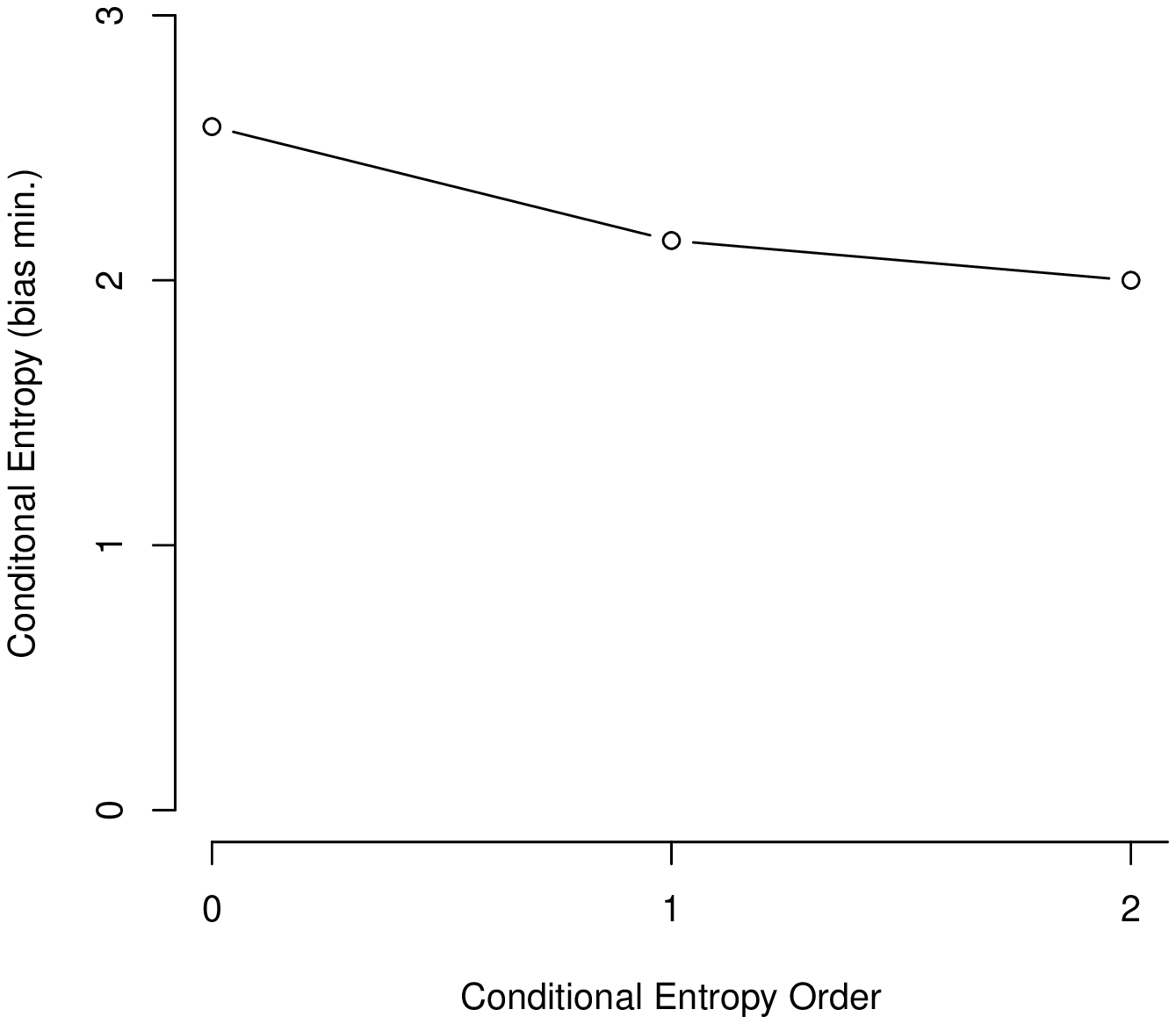}\\
    \small{Bottlenose dolphins (\emph{Tursiops truncatus})}&\small{Humpback Whales (\emph{Megaptera novaeangliae})}\\
	 \includegraphics[height=2.5in, width=2.5in]{skylarkinfograph.eps}&
    \includegraphics[height=2.5in, width=2.5in]{starlinginfograph.eps}\\ 	 
	    \small{European skylarks (\emph{Alauda arvensis L.})}&\small{European starlings (\emph{Sturnus vulgaris})}\\
	 \includegraphics[height=2.5in, width=2.5in]{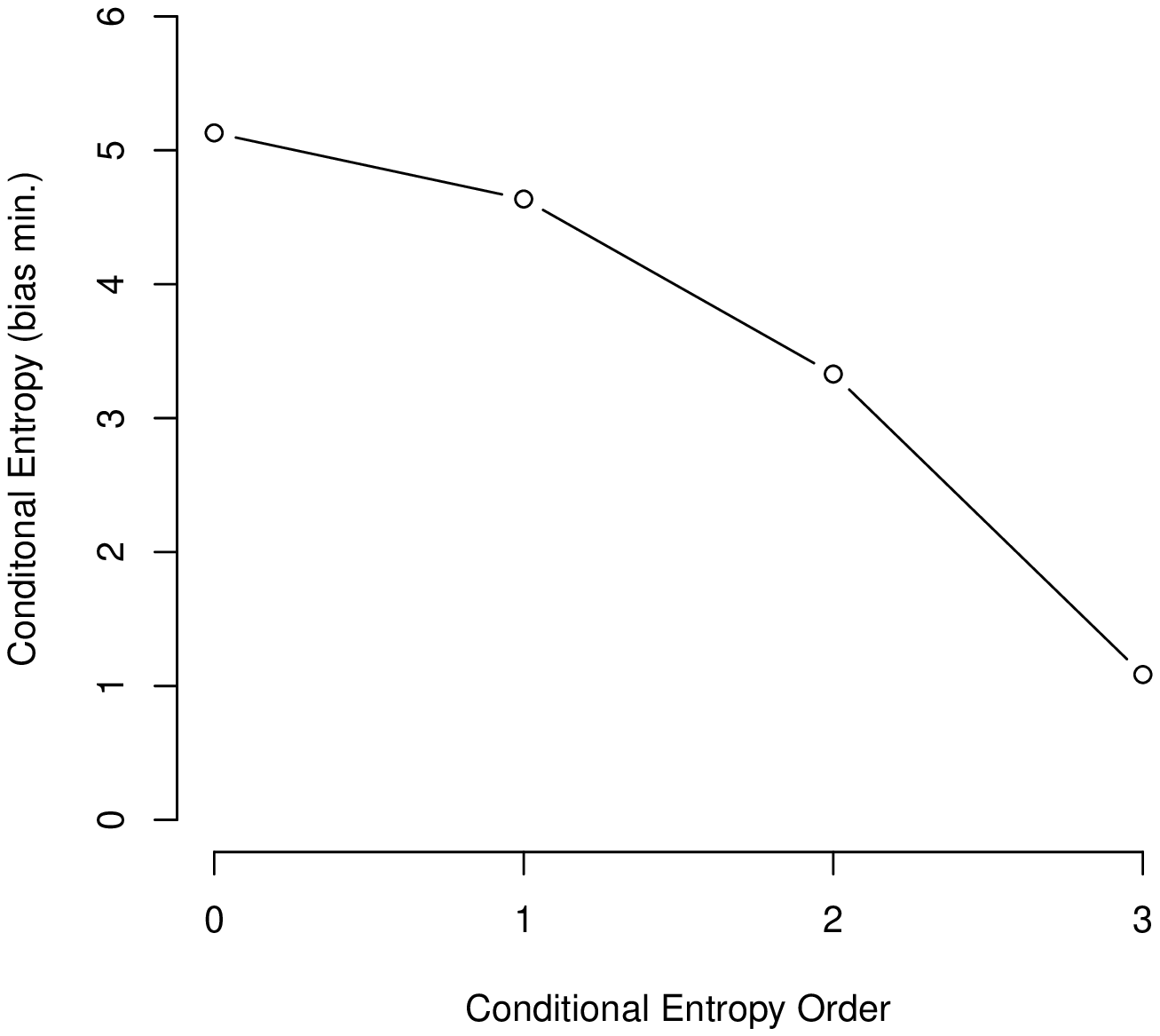}&
    \includegraphics[height=2.5in, width=2.5in]{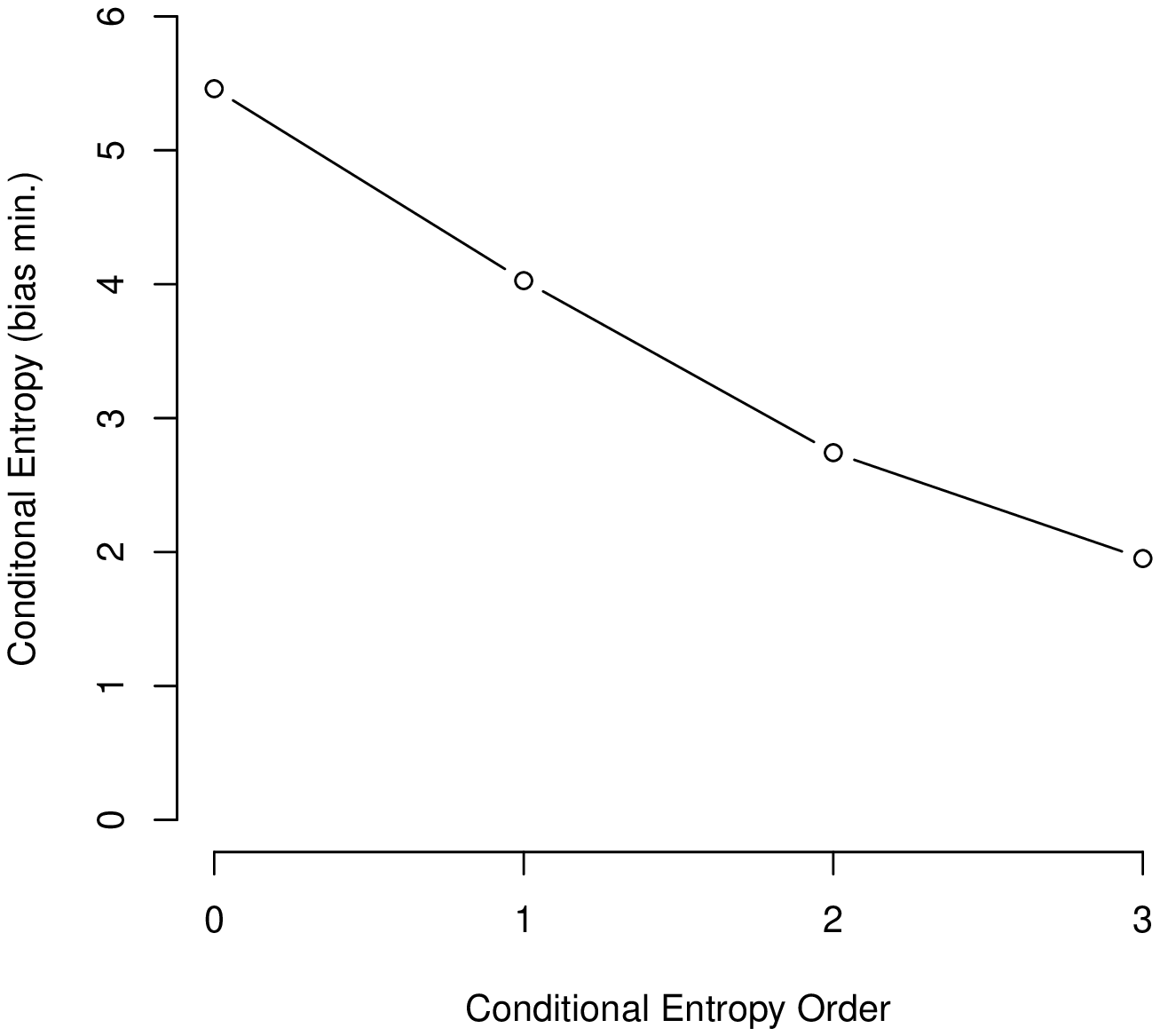}\\
	    \small{Wood thrushes (\emph{Hylocichla mustelina})}&\small{Robins (\emph{Turdus migratorius})}\\
\end{tabular}
\caption{Information graphs of animal communication conditional entropies for the species analyzed in this paper.}
\label{allinfograph}
\end{figure}

Here we use the data from these papers to reproduce graphically the information graphs for the communications of each species (Figure \ref{allinfograph}) as well as to show the conditional entropy for the first three orders, correct the conditional entropy for bias, and estimate the minimum and maximum size of the animal N-gram repertoires given the bias corrected entropy values.

First, we will represent the minimum bias corrected conditional entropies as information graphs from order 0, $\log M$ for the number of individual symbols, to the third order. Only the humpback whale data stops at the second order due to a lack of data on the third order entropy.

\clearpage

As can be seen in Figure \ref{allinfograph}, several species show a dramatic drop after the first or second order of entropy. For a basic comparison, the information graph for written English letters (\cite{shannonenglish, dolphincond}) shows a much more gradual decline and thus less repetition. Once again, it is difficult to make a definite interpretation of the order of the process with sample sizes that are not as large as or are barely larger than $M^2$, especially with the large song type repertoire of birds.

In analyzing the data from the species and estimating repertoires it is essential to define sample sizes and correct for bias. In Table \ref{paperdatatable}, the basic data from the papers is shown. One key issue to resolve is which sample size to use at each order. Sample sizes for higher order N-grams can be reduced if there are multiple discrete sequences. For example, if there are 500 individual symbols in a dataset, yet these are broken into 25 discrete sequences, the first order sample size is 500 while the second order must be 475 since there is no overlap with the end of one sequence and the beginning of another. This information was not always available but for dolphins, humpback whales, and starlings, this methodology was used to calculate $S_2$ and $S_3$.

In Table \ref{biastable}, the minimum bias and maximum bias for each species are given. For the maximum bias, there were exceptions where the symbol size dictated by $H$ was so large that the bias correction would cause the conditional entropy to exceed the value of the previous order. In this case the bias was limited to the maximum possible value--that which would make the conditional entropy at this order (usually the third order) equal to that of the second order.

In Tables \ref{correntropytable} and \ref{repertoiretable}, the final estimates for the bias corrected conditional entropies and the derived repertoire sizes are given.

\begin{figure}
\centering
\includegraphics[height=2.5in, width=2.5in]{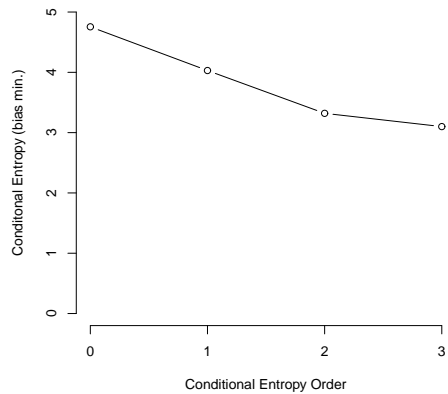}
\caption{Information graph of written English letters based on \cite{shannonenglish, dolphincond}. The smaller negative slope as compared with bird song information graphs is evident as first shown in \cite{chick1}.}
\end{figure}
\clearpage
\begin{table}
\centering
\small
\begin{tabular}{|c|c|c|c|c|c|c|c|c|}
\hline 
Species Name&Reference&$M$&$S$&$S_2$&$S_3$&$H$&$H_2$&$H_3$\\
\hline 
\emph{Tursiops truncatus}&\cite{dolphincond}&27&493&346&346&1.92&1.15&0.56\\
\hline
\emph{Megaptera novaeangliae}&\cite{whalecond}&6&202&195&N/A&2.15&2&N/A\\
\hline 
\emph{Alauda arvensis L.}&\cite{skylark}&170&10000&10000&10000&7.05&1&0.29\\
\hline 
\emph{Sturnus vulgaris}&\cite{starling}&105&4811&4691&4691&6.03&1.47&0.81\\
\hline 
\emph{Hylocichla mustelina}&\cite{lemondobson}&35&777&777&777&4.64&3.33&1.09\\
\hline 
\emph{Turdus migratorius}&\cite{lemondobson}&44&2700&2700&2700&4.03&2.74&1.95\\
\hline 
\end{tabular}
\caption{The basic data on the information theory of animal communication from the species analyzed. $M$ is the number of base symbols (songs, whistles, cries, etc.), $S$ is the sample size of symbols analyzed, $S_2$ is the estimate (where available) of the number of 2-grams measured, $S_3$ is the estimate (where available) of the number of 3-grams measured, $H$, $H_2$, and $H_3$ are the first, second, and third order conditional entropies respectively.}
\label{paperdatatable}
\end{table}

\begin{table}
\centering
\small
\begin{tabular}{|c|c|c|c|c|c|c|}
\hline 
&\multicolumn{3}{|c|}{Bias Min}&\multicolumn{3}{|c|}{Bias Max}\\
\hline 
Species Name&$H$&$H(X,Y)$&$H(X,Y,Z)$&$H$&$H(X,Y)$&$H(X,Y,Z)$\\
\hline 
\emph{Tursiops truncatus}&0.04&0.01&0.01&0.04&0.03&0.2\\
\hline
\emph{Megaptera novaeangliae}&0.02&0.06&N/A&0.02&0.07&N/A\\
\hline 
\emph{Alauda arvensis L.}&0.01&0&0&0.01&1.26&1.96*\\
\hline 
\emph{Sturnus vulgaris}&0.02&0.0&0.0&0.02&0.66&1.3\\
\hline 
\emph{Hylocichla mustelina}&0.03&0.09&0.01&0.03&0.57&2.78*\\
\hline 
\emph{Turdus migratorius}&0.01&0.01&0.02&0.01&0.07&0.85*\\
\hline 
\end{tabular}
\caption{The biases, minimum and maximum, calculated for the joint entropies of orders 1-3 according to the paper data. Values with asterisks indicate where the maximum bias assumption correction would have exceeded the previous order entropy and therefore the maximum bias is limited to the difference between the bias-corrected previous order entropy and the original entropy estimate.}
\label{biastable}
\end{table}

\begin{table}
\centering
\small
\begin{tabular}{|c|c|c|c|c|c|c|}
\hline 
&\multicolumn{3}{|c|}{Bias Min}&\multicolumn{3}{|c|}{Bias Max}\\
\hline 
Species Name&$H$&$H_2$&$H_3$&$H$&$H_2$&$H_3$\\
\hline 
\emph{Tursiops truncatus}&1.96&1.12&0.56&1.96&1.14&0.73\\
\hline
\emph{Megaptera novaeangliae}&2.17&2.04&N/A&2.17&2.05&N/A\\
\hline 
\emph{Alauda arvensis L.}&7.06&0.99&0.29&7.06&2.25&2.25*\\
\hline 
\emph{Sturnus vulgaris}&6.05&1.46&0.81&6.05&2.11&2.09\\
\hline 
\emph{Hylocichla mustelina}&4.67&3.39&1.00&4.67&3.87&3.87*\\
\hline 
\emph{Turdus migratorius}&4.04&2.74&1.96&4.04&2.8&2.8*\\
\hline 
\end{tabular}
\caption{The corrected conditional entropies, minimum and maximum, calculated for the conditional entropies of orders 1-3 according to the paper data and values in Tables \ref{paperdatatable} and \ref{biastable}. Values with asterisks indicate where the maximum bias assumption correction would have exceeded the previous order entropy and therefore the maximum bias is limited at the bias-corrected previous order entropy.}
\label{correntropytable}
\end{table}

\clearpage

\begin{table}
\centering
\small
\begin{tabular}{|c|c|c|c|c|c|c|c|c|}
\hline 
&\multicolumn{4}{|c|}{Bias Min}&\multicolumn{4}{|c|}{Bias Max}\\
\hline 
Species Name&1-gram&2-gram&3-gram&Total&1-gram&2-gram&3-gram&Total\\
\hline 
\emph{Tursiops truncatus}&27&5&4&36&27&5&5&37\\
\hline
\emph{Megaptera novaeangliae}&6&17&N/A&23&6&18&N/A&24\\
\hline 
\emph{Alauda arvensis L.}&170&4&2&176&170&23&108*&301\\
\hline 
\emph{Sturnus vulgaris}&105&8&6&119&105&19&78&202\\
\hline 
\emph{Hylocichla mustelina}&35&110&8&153&35&214&3126*&3375\\
\hline 
\emph{Turdus migratorius}&44&45&59&148&44&49&338*&431\\
\hline 
\end{tabular}
\caption{Estimates of total repertoire sizes for 1-gram, 2-gram, and 3-gram, minimum and maximum, for each species based on the bias corrected conditional entropies.}
\label{repertoiretable}
\end{table}

From these tables, especially Table \ref{repertoiretable}, several things seem clear. First, for almost all of the species given, the bulk of their N-gram repertoire lies within the 1-gram individual symbols. The largest exceptions, for both the maximized and minimized bias, seem to be the wood thrush and robins. There could be exceptions, however. For example, in \cite{dolphincond}, the authors used only those dolphin whistles that occurred at least twice for entropy calculations giving an $M=27$. There were a total of 102 distinct whistles detected, 75 only once, so adding these would give a total repertoire for the dolphins of 112 for the maximum bias and 111 for the minimum bias. 

Clearly, we have a more accurate idea of total repertoire with those animals where the repertoire size differs very little from the maximum or minimum bias assumptions. These are dolphins, humpback whales, and European starlings. The other bird species have a large number of song types. This huge symbol size causes a large swing between the estimates for minimum and maximum bias. In these cases, the minimum bias estimate is more representative since the number of possible N-grams that first-order entropy would imply is enormous with such  a large symbol set. In the end, the best way to accurately measure the repertoire sizes, particularly for dolphins and humpback whales, is to make a much larger measurement of sequences with $S$ in the thousands.

\section{Other Repertoire Counting Methods and Simulation}

As stated in the introduction, apart from the information theory perspective, repertoire size has often been investigated using sampling methods such as curve-fitting and capture-recapture. These methods can be used if song bout data is available to predict repertoire size, their accuracy increasing with the number of samples. In order to compare the method developed in this paper with actual data and these two methods, a program was created that synthesized an arbitrary signal with a predefined entropy of the first, second, and third order.

\begin{figure}
\centering
 \begin{tabular}{cc}
	 \includegraphics[height=2.5in, width=2.5in]{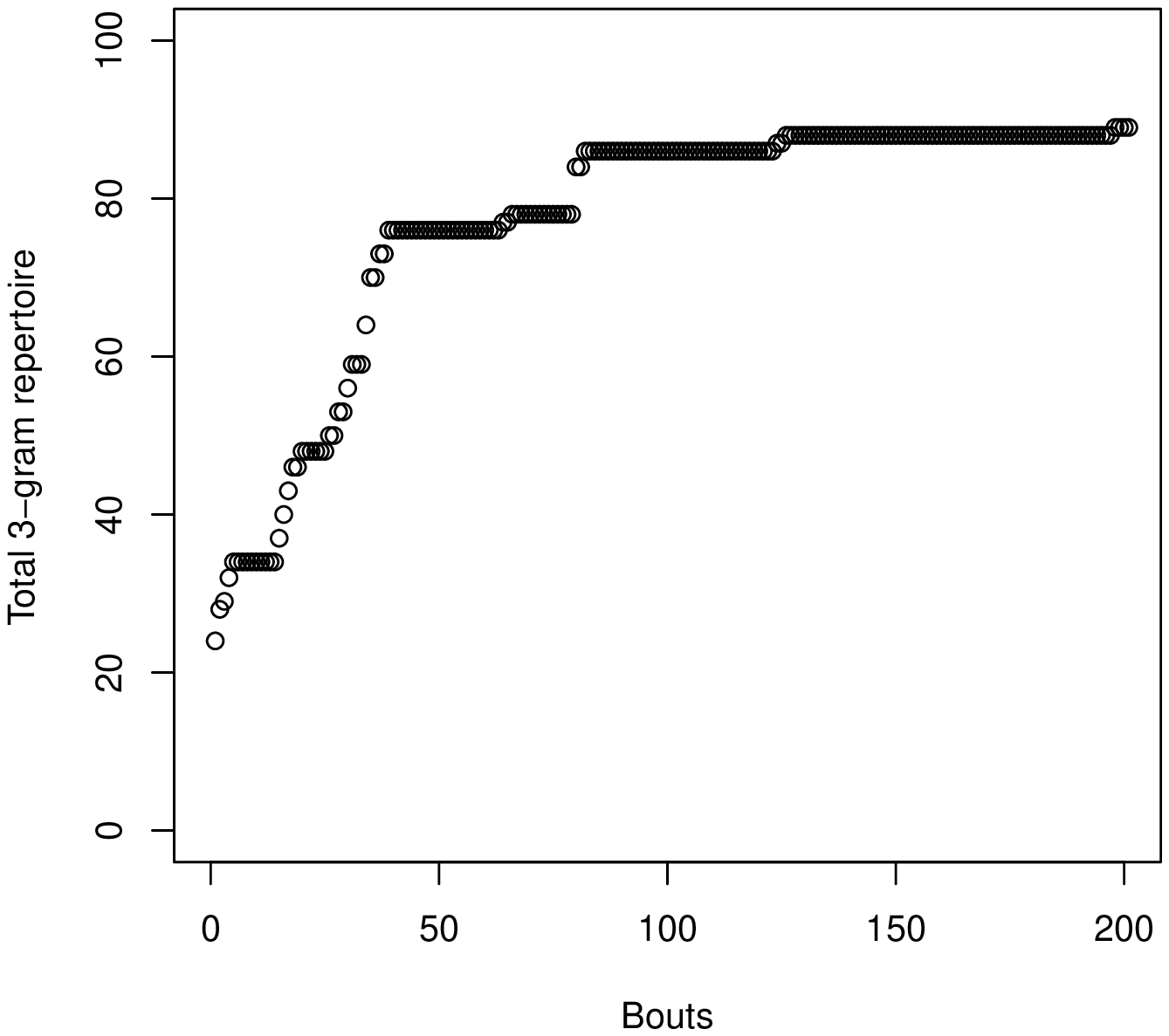}&
    \includegraphics[height=2.5in, width=2.5in]{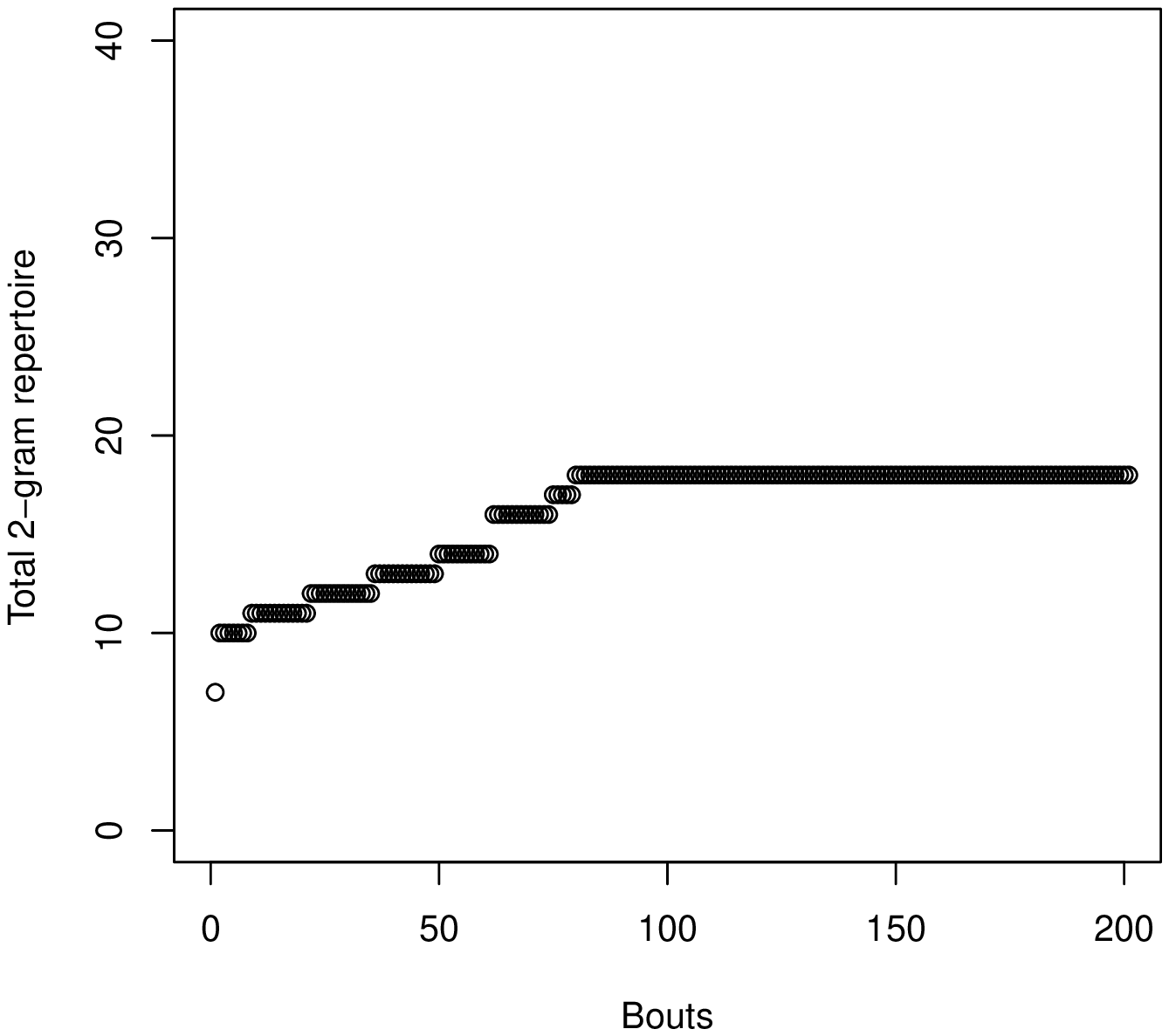}\\
    \small{Bottlenose dolphins 3-grams}&\small{Humpback whales 2-grams}\\
\end{tabular}
\caption{Exponential distribution of repertoire growth over time for bottlenose dolphin 3-grams and humpback whale 2-grams. Based on simulated sequences of 20,000 symbols with repertoire measured in bouts of 100 symbols for dolphins and a sequence of 2,000 symbols with bouts of 10 symbols for humpback whales.}
\label{dolphinwhalerep}
\end{figure}

Using this program, the number of N-grams was compared with the estimates using the entropy method for dolphins and humpback whales. For dolphins and whales respectively, 20,000 symbol and 2,000 symbol sequences with matching conditional entropies were created and the number of N-grams from 1 to 3 were counted. Since the samples were so large, neither curve-fitting nor capture-recapture had an issue finding the total repertoire size since the exponential distribution of the total number of symbols (see Figure \ref{dolphinwhalerep}) reaches as asymptote. Part of the reason for the rapid symbol acquisition may be that the sequences, despite having the requisite entropy properties, were relatively stationary which is not always the case for real languages. For dolphins and whales, the charts in the tables were created by sampling new symbols in song `bouts' of 100 and 10 symbols respectively. 

For the humpback whales, the total number of simulated 2-grams exactly matched the prediction of a repertoire size of 18. This would seem to confirm the validity of the method. The dolphin story was more complex. With dolphins, the total number of simulated N-grams, exceeded the values estimated by the entropy estimations in all cases, however, the details tell a more complex story. While the repertoire is large in terms of N-grams, the frequency is very concentrated amongst the top N-grams. The top 5 2-grams and 3-grams are 78\% and 63\% of all 2-grams (total: 46) and 3-grams (total: 89) respectively. Many of the 2-grams and 3-grams occurred only once in the 20,000 symbol sequence. While the bias in the dolphins is greater due to the relatively small sample size compared to the number of symbols, the repertoire exceeded even the maximum bias estimates for both 2-grams and 3-grams.

Therefore, we can conclude one major strength, but limitation, of the use of conditional entropy to measure the N-gram repertoire. For small repertoires, like the whales, it seems they can accurately estimate repertoires for small combinations such as 2-grams. For more complex repertoires, they seem to accurately measure the size of the most frequently used N-grams in the repertoire to give a reasonable estimate of the most functionally used N-grams. As a limit, however, conditional entropies can seriously undercount rare N-grams since their relatively small probabilities contribute to the calculations of entropy only weakly.

If collecting the entire size of the repertoire, ignoring the weighted heterogeneity of the symbols, is desired and samples are available, both curve-fitting and capture-recapture create a more detailed picture since they can pick up the rare occurrences, however, they do not give the same information about the relative skewed nature of the distribution of symbols the entropy method can provide.

\section{Conclusions}

Animal communication analyses through information theory have been useful, and while they cannot answer all questions regarding the intent or possible meaning of such communications, they have shown beyond a doubt that animal communication can have a complex structure that goes beyond random sounds or even the structure of a first-order Markov process.

However, entropy based analyses alone hold only descriptive power. A logical next step from observing and measuring communications complexity should be determining how to use that complexity to search for communications structures that can help understand animal behavior. The methods outlined in this paper assist in this effort by giving researchers a baseline to investigate further regarding 2-gram or 3-gram call sequences. In particular, the size of the most frequent, and possibly functional, repertoire is clearly enumerated using information theory methods. Similar to work by Getner on starlings \cite{starling}, these analyses can reveal that single songs or cries are poor substitutes for communication outside the complete pattern. Assumptions of uniform probabilities for the repertoire are almost always wrong and plain measures of repertoire size cannot reflect this as well as entropy values.

While the information theory methods are weaker in finding the exact repertoire size compared to count based methods such as curve-fitting and capture-recapture, these methods offer an improved understanding of the relationships that develop the syntax of the communication. The basic order of communication, the clustering of ``vocabulary'', and other detailed features cannot been understood just by comparing repertoire sizes over time and across species. The importance of understanding syntax in this matter has been frequently raised such as in \cite{yip} where it is recommended that more experiments be carried out to ascertain if other species have phonological recognition similar to phonemes in human speech. 

It has long been known that auditory recognition abilities exist in a wide group of species from 2-gram alarm calls in putty monkeys (\emph{Cercopithecus nictitans}) \cite{arnold1, arnold2} to pitch differentiation by moths \cite{turner,turner2}. How and why these abilities could possibly exist in disparate species such as birds and cetaceans while possibly absent in some more closely related primates is a key question. Is this a frequent evolutionary adaptation that can appear in almost any species or do the most elaborate and complex communications, such as with dolphins, require high intelligence \cite{lilly}?

Just like word length analyses in human language use syllables as the base unit \cite{altmann}, we may possibly look at the average, or most frequent, length of N-grams of communication in animals to gauge the depth and complexity of their communications. In this way, it is the author's hope that information theory analyses can help peel back the layers of complexity to show how closely such animal communication matches--or is distinct from--human language. 

\acknowledgements{Acknowledgments}

I would like to thank Laurance Doyle for help gathering and understanding data from past papers. I would also like to thank the anonymous referees for much helpful feedback.

\acknowledgements{Conflicts of Interest}
The author has no conflicts of interest.

\bibliographystyle{mdpi}
\makeatletter
\renewcommand\@biblabel[1]{#1. }
\makeatother

\end{document}